\documentstyle[12pt,epsf]{article}

\expandafter\ifx\csname amssym12.def\endcsname\relax \else\endinput\fi
\expandafter\edef\csname amssym12.def\endcsname{%
       \catcode`\noexpand\@=\the\catcode`\@\space}
\catcode`\@=11

\def\undefine#1{\let#1\undefined}
\def\newsymbol#1#2#3#4#5{\let\next@\relax
 \ifnum#2=\@ne\let\next@\msafam@\else
 \ifnum#2=\tw@\let\next@\msbfam@\fi\fi
 \mathchardef#1="#3\next@#4#5}
\def\mathhexbox@#1#2#3{\relax
 \ifmmode\mathpalette{}{\m@th\mathchar"#1#2#3}%
 \else\leavevmode\hbox{$\m@th\mathchar"#1#2#3$}\fi}
\def\hexnumber@#1{\ifcase#1 0\or 1\or 2\or 3\or 4\or 5\or 6\or 7\or 8\or
 9\or A\or B\or C\or D\or E\or F\fi}

\font\tenmsa=msam10 scaled\magstep1
\font\sevenmsa=msam7 scaled\magstep1
\font\fivemsa=msam5 scaled\magstep1
\newfam\msafam
\textfont\msafam=\tenmsa
\scriptfont\msafam=\sevenmsa
\scriptscriptfont\msafam=\fivemsa
\edef\msafam@{\hexnumber@\msafam}
\mathchardef\dabar@"0\msafam@39
\def\dashrightarrow{\mathrel{\dabar@\dabar@\mathchar"0\msafam@4B}}
\def\dashleftarrow{\mathrel{\mathchar"0\msafam@4C\dabar@\dabar@}}

\def\ulcorner{\delimiter"4\msafam@70\msafam@70 }
\def\urcorner{\delimiter"5\msafam@71\msafam@71 }
\def\llcorner{\delimiter"4\msafam@78\msafam@78 }
\def\lrcorner{\delimiter"5\msafam@79\msafam@79 }
\def\yen{{\mathhexbox@\msafam@55 }}
\def\checkmark{{\mathhexbox@\msafam@58 }}
\def\circledR{{\mathhexbox@\msafam@72 }}
\def\maltese{{\mathhexbox@\msafam@7A }}

\font\tenmsb=msbm10 scaled\magstep1
\font\sevenmsb=msbm7 scaled\magstep1
\font\fivemsb=msbm5 scaled\magstep1
\newfam\msbfam
\textfont\msbfam=\tenmsb
\scriptfont\msbfam=\sevenmsb
\scriptscriptfont\msbfam=\fivemsb
\edef\msbfam@{\hexnumber@\msbfam}

\def\widehat#1{\setbox\z@\hbox{$\m@th#1$}%
 \ifdim\wd\z@>\tw@ em\mathaccent"0\msbfam@5B{#1}%
 \else\mathaccent"0362{#1}\fi}
\def\widetilde#1{\setbox\z@\hbox{$\m@th#1$}%
 \ifdim\wd\z@>\tw@ em\mathaccent"0\msbfam@5D{#1}%
 \else\mathaccent"0365{#1}\fi}
\font\teneufm=eufm10 scaled\magstep1
\font\seveneufm=eufm7 scaled\magstep1
\font\fiveeufm=eufm5 scaled\magstep1
\newfam\eufmfam
\textfont\eufmfam=\teneufm
\scriptfont\eufmfam=\seveneufm
\scriptscriptfont\eufmfam=\fiveeufm

\csname amssym.def\endcsname

\newif{\ifcomentarios}
\comentariosfalse

\renewcommand{\theequation}{\thesection.\arabic{equation}}

\newcommand{\zerarcounters}
{
\setcounter{equation}{0}
\setcounter{theorem}{0}
}

\newcommand{\be}{\begin{equation}}
\newcommand{\ee}{\end{equation}}
\newcommand{\bma}{\begin{displaymath}}
\newcommand{\ema}{\end{displaymath}}
\newcommand{\bc}{\begin{center}}
\newcommand{\ec}{\end{center}}

\newcommand{\text}{\rm}

\newcommand{\uflex}
{{\scriptstyle {\raise 9pt\hbox{$\backslash$}\,\!\!\!\!\!\Bigg\vert}}}

\newcommand{\ncm}{\newcommand}

\ncm{\rncm}{\renewcommand}
\ncm{\id}{{\bf 1}}
\ncm{\beq}{\begin{equation}}
\ncm{\eeq}{\end{equation}}
\ncm{\ba}{\begin{array}}
\ncm{\bea}{\begin{eqnarray}}
\ncm{\beanon}{\begin{eqnarray*}}
\ncm{\ea}{\end{array}}
\ncm{\eea}{\end{eqnarray}}
\ncm{\eeanon}{\end{eqnarray*}}
\ncm{\fns}{\footnotesize}
\ncm{\setc}[1]{\setcounter{equation}{#1}}
\newcounter{eqnr}
\newenvironment{eqnarrayabc}{\stepcounter{equation}
  \setcounter{eqnr}{\value{equation}}\setc{0}
  \rncm{\theequation}{\thesection.\arabic{eqnr}\alph{equation}}
  \begin{eqnarray}}{\end{eqnarray}\setc{\value{eqnr}}}
\ncm{\eqboxabc}[3]{\newline\parbox[t]{1.5cm}{#1}\hfill
  \parbox[b]{12cm}{\begin{eqnarray*} #3\end{eqnarray*}}\hfill
   \parbox[b]{1.5cm}{\vspace{-0.0cm}\begin{eqnarrayabc}#2\end{eqnarrayabc}}\newline}
\ncm{\eqbox}[2]{\newline\parbox{1.5cm}{#1}\hfill
  \parbox{12cm}{\beanon #2\eeanon}\hfill
  \parbox{1cm}{\bea\eea}\newline}
\ncm{\nr}[1]{\parbox{1cm}{\begin{eqnarrayabc}#1\end{eqnarrayabc}}\\}

\ncm{\kal}[1]{\mbox{$\cal #1 $}}
\ncm{\mrk}[1]{\!\!\! #1 \!\!\!} 
\ncm{\qed}{\hspace*{0.4cm}\rule{0.24cm}{0.24cm}}  
\ncm{\mbold}[1]{\mbox{\boldmath $ #1 $}}   
\ncm{\bm}{\mbold}
\ncm{\str}{\stackrel}
\ncm{\sub}{\subset}
\ncm{\e}{\varepsilon}
\ncm{\ka}{\kappa}
\ncm{\inputc}[1]{\begin{center}\input{#1}\end{center}}
\ncm{\lto}{\longrightarrow}
\ncm{\x}{\times}
\ncm{\bmm}{\bm{\cal M}}
\ncm{\cp}{{\bf P}}    
\ncm{\bfp}{{\bf P}}
\ncm{\bmi}{\bm{i}}
\ncm{\bmom}{\bm{\om}}
\ncm{\bmOm}{\bm{\Om}}
\ncm{\res}{\restriction}
\ncm{\bmL}{\bm{\cal L}}
\ncm{\bmell}{\bm{\ell}}
\ncm{\bmE}{\bm{\cal E}}
\ncm{\bme}{\bm{e}}
\ncm{\bmpi}{\bm{\pi}}
\ncm{\bmr}{\bm{r}}
\ncm{\bmsigma}{\bm{\sigma}}
\ncm{\wt}{\widetilde}

\newcommand{\beaa}{\begin{eqnarray}}
\newcommand{\eeaa}{\end{eqnarray}}

\begin{document}
\input{epsf.tex}

\author{{\bf Oscar Bolina}\thanks{Supported by FAPESP under grant
97/14430-2. {\bf E-mail:} 
bolina@lobata.math.ucdavis.edu} \\
Department of Mathematics\\
University of California, Davis\\
Davis, CA 95616-8633, USA\\
\\
{\bf L. H. A. Monteiro}\thanks{Supported by FAPESP 
under grant 97/01645-0. {\bf E-mail:} luizm@lac.usp.br}\\ 
Departamento de Engenharia Eletr\^onica \\
Universidade de S\~ao Paulo \\
05508-900 S\~ao Paulo, Brasil \\
}
\title{\vspace{-1in}
{\bf A Note on Eye Movement}}
\date{}
\maketitle
\begin{abstract}
\noindent
In a simplified fashion, the motion of the eyeball in its orbit 
consists of rotations around a fixed point. Therefore, this motion 
can be described in terms of the Euler's angles of rigid body
dynamics. However, there is a physiological constraint in the motion 
of the eye which reduces to two its degrees of freedom, so that one 
of Euler's angles is not an independent variable. This paper reviews 
the basic features of the kinematics of the eye and the laws governing 
its motion.

\hfill \break
\end{abstract}



\section{Introduction}
\zerarcounters

The kinematics of the human eye has been extensively investigated 
in physics and medicine \cite{{H},{S},{He},{L},{K},{Ha}}. From the 
wealth of material available we have selected a few topics of 
more direct interest to physicists concerning with the
description of rotational motion of the eye and its biological
constraints.
\newline
In the simplified model of the eye, the organ is considered to be 
a spherical surface held by six muscles in a bony cavity known as
{\it orbit} or {\it eye socket}. In the relaxed normal eye, the 
pressure of internal fluids maintains the radius of curvature of 
the eye at a constant value of approximately 12 mm. Since the actual 
size of the eyeball is in fact irrelevant to our analysis, as soon 
as it is constant, we assume the eyeball to be a spherical surface
of unit radius. 
\newline
We also disregard any translational motion of the eye with respect 
to the head, and consider only the rotary motions of the eyeball
in the orbital socket. To a good approximation the center of rotation 
is fixed and coincides with the center of the eyeball {\it O}. 
\newline
The basic rotations of the eye consist of its motions in looking
from side to side along the horizontal meridian (a rotation around
a vertical axis), and up and down (a rotation around a horizontal 
axis). These two motions are called {\it cardinals}.
\newline
There is a third possible motion called {\it torsion}, which is 
a rotation of the eye to compensate for an eventual tilting of the 
head while keeping the same direction of vision. Torsion also
occurs when the pupil moves forcibly up or down, approaching the 
nose \cite{DE}. 
\newline
Instead of the cardinal angles just mentioned, we prefer to 
describe the motion of the eyeball in terms of Euler's angles.


\section{Euler's Description of Motion}
\zerarcounters

Suppose one holds the head upright and looks directly toward a
distant target object horizontally straight in front. The line
extending from {\it O} to the fixation point is called the
{\it visual axis}, and this position defines the {\it primary
position} of the visual axis, to which all other will be referred
(See \cite{O} for some remarks on this). In changing the direction of
the visual axis as the target object moves, the eye rotates around the
point {\it O}.
\newline
For the description of the rotational motion we set up a system of 
axis {\it OABC} attached to the eyeball and moving with it. The
position of the moving axes relative to a head-fixed system 
{\it OXYZ} is determined by Euler's angles. Assume initially the
two systems coincide, and take the primary position of the visual
axis to be the horizontal direction {\it OZ} or {\it OC} (The eye
is looking to the reader), as shown in Fig. 1. 
\newline
Starting from the primary position a general displacement of the 
eyeball can be effected by the following sequence of rotations
(all of which performed in anti--clockwise direction). The
eyeball is first turned around the axis {\it OZ} through an angle
$\psi$. This rotation brings the axes {\it OX} and {\it OY} to the 
positions {\it OX'} and {\it OY'}. A second rotation around the axis 
{\it OY'} through an angle $\theta$ moves {\it OZ} to the position 
{\it OC} and {\it OX'} to the position {\it OA'} both on the plane
{\it OZX'}. A final rotation around {\it OC} through an angle $\phi$ 
brings {\it OA'} to {\it OA} and {\it OY'} to {\it OB} both on the 
same plane {\it OA'Y'} \cite{Scar}.
\newline
In this representation, the {\it torsion motion} of the eyeball  
is simply the third rotation around the visual axis by an
angle $\phi$. 


\section{Donders' Law}
\zerarcounters

In the movements of the eye there is a physiological constraint 
that limits the motion of the eyeball around the visual axis.
\newline
According to Donders' law \cite{DE}:
\begin{itemize}
\item[{}] {\it Every secondary position of the eyeball is associated
with a definite and constant amount of torsion, no matter how the
position be reached.}
\end{itemize}
This means that the eye needs only two parameters to be fully described,
the value of the angle of torsion $\phi$ being fixed for given $\psi$ 
and $\theta$ \cite{C}. 
\newline
This natural reduction of the degrees of freedom is connected with the
fact that when a target object is viewed with the head in a certain
position relative to it, the image of the object must be formed on 
the same region of the retina whenever the gaze is directed to the 
same point of the object, no matter what the movements the eye
might have made between two gazes of the same object \cite{D}.
\newline
It is easy to see that after the rotations by the angles $\psi$ and
$\theta$, the eyeball at the position {\it C} will have turned by an
amount $\psi$ relative to the object viewed in comparison with its
relative position at the primary position {\it Z}. In order that the
gaze affect the same elements of the retina the third rotation around
{\it OC} by an amount $\phi$ must restore the initial relative position
of the eyeball and the object. This is accomplished by performing a
rotation $\phi=-\psi$ around {\it OC}. 


\section{Listing's Law}
\zerarcounters

Listing explained how the position of the eye could be described
by only two independent quantities. According to Listing's law 
for the motion of the eyeball \cite{C}: 
\begin{itemize}
\item[{}] {\it Any movement of the eye from the primary position to 
any other secondary position is equivalent to a single rotation 
around an axis perpendicular both to OZ and OC through an angle ZOC.}
\end{itemize}
In Fig. 1 this axis of rotation is the axis {\it OY'}, making an angle
$\psi$ (the meridional angle) with the vertical plane in the primary
position, and the angle of rotation is $\theta$ (also called the
eccentricity).
\newline
This law does not specifies how the eye actually moves from the 
primary position to any secondary position. It is not even necessary
that the motion be effected in this way, but the result must be the
same whatever the path the gaze may have taken to reach that final
position \cite{Ho} (Compare with Euler's theorem: The general
displacement of a rigid body with one point fixed is a rotation 
around some axis).
\newline
In these terms, Donders' law appears as a statement that the position 
of the eye as a whole is uniquely determined by the direction of the
visual axis.
\newline
The constraint on the movement of the eye expressed by Listing's law 
can not be of mechanical nature only, but may also have a neurological 
component, since it has been observed that the eyes do not obey Listing's
law during sleep \cite{Ho}.


\section{Torsion and False Torsion}
\zerarcounters

In the motion of the eye away from the primary position in the manner 
prescribed by Listing's law, which is by a rotation around {\it
OY'} through an angle $\theta$, the vertical meridian plane of the
eye in the primary position {\it OYZ} traces out the surface of a
cone of vertex {\it O} and semi-angle $\psi$ formed by the axes
{\it OY} and {\it OY''} and bisected by the axis of rotation {\it
OY'}, as shown in Fig. 2. In its rotation the axis {\it OY} moves
to its new position {\it Oy}. The axes {\it OY}, {\it OY''} and 
{\it Oy} meet the unit sphere at the points {\it P,Q, R}
respectively. The straight line {\it PQ} cuts the axis {\it OY'} at
{\it T} defining the lengths $PT=\sin\psi$ and $OT=\cos\psi$.  
\newline
This motion is associated with a definite inclination (also called
a tilt or {\it torsion}) of the vertical meridian plane in its
secondary position {\it OyC} with respect to the vertical plane
{\it OYC} passing through {\it OY} and the visual axis in the
secondary position \cite{A}.
\newline
The inclination depends on the meridional and eccentricity angles
in the following way. In Fig. 2, the vertical plane {\it OYC} is 
normal to the plane {\it OTR}, and cuts the line {\it TR} at {\it S}.
The angle of inclination $\rho$ is the angle between {\it ORS} in the
plane {\it ORT}. From the triangle {\it OTR} we have \cite{DE}
\[
\tan(\psi-\rho)=\frac{ST}{OT},
\]
where $ST=PT \cos\theta$ and $OT=\cos\psi$. Thus
$\tan(\psi-\rho)=\tan\psi \cos\theta$. Solving for $\tan \rho$ we get
(Check it \cite{DE} and compare with \cite{{C},{A}})
\begin{equation}
\tan \rho=\frac{\tan \psi(1-\cos \theta)}{1+\tan^{2}\psi \cos\theta}.
\end{equation}
In our use of Euler' angles, the torsional rotation around {\it OC}
by $-\psi$ was intended to bring the eyeball to the position it would
assume under Listing's law. Since the final orientation of the eye
depends on the way rotations are carried out, had we chosen another 
set of axis to describe the motion of the eye we would have to perform
a torsion motion of different magnitude. Because of this, the torsional
motion required to be performed in order to bring the eyeball to the
position specified by Listing's law when using any other system of axis
is often called a {\it false torsion} \cite{C}.


\newpage 
\begin{figure}
\centerline{
\epsfbox{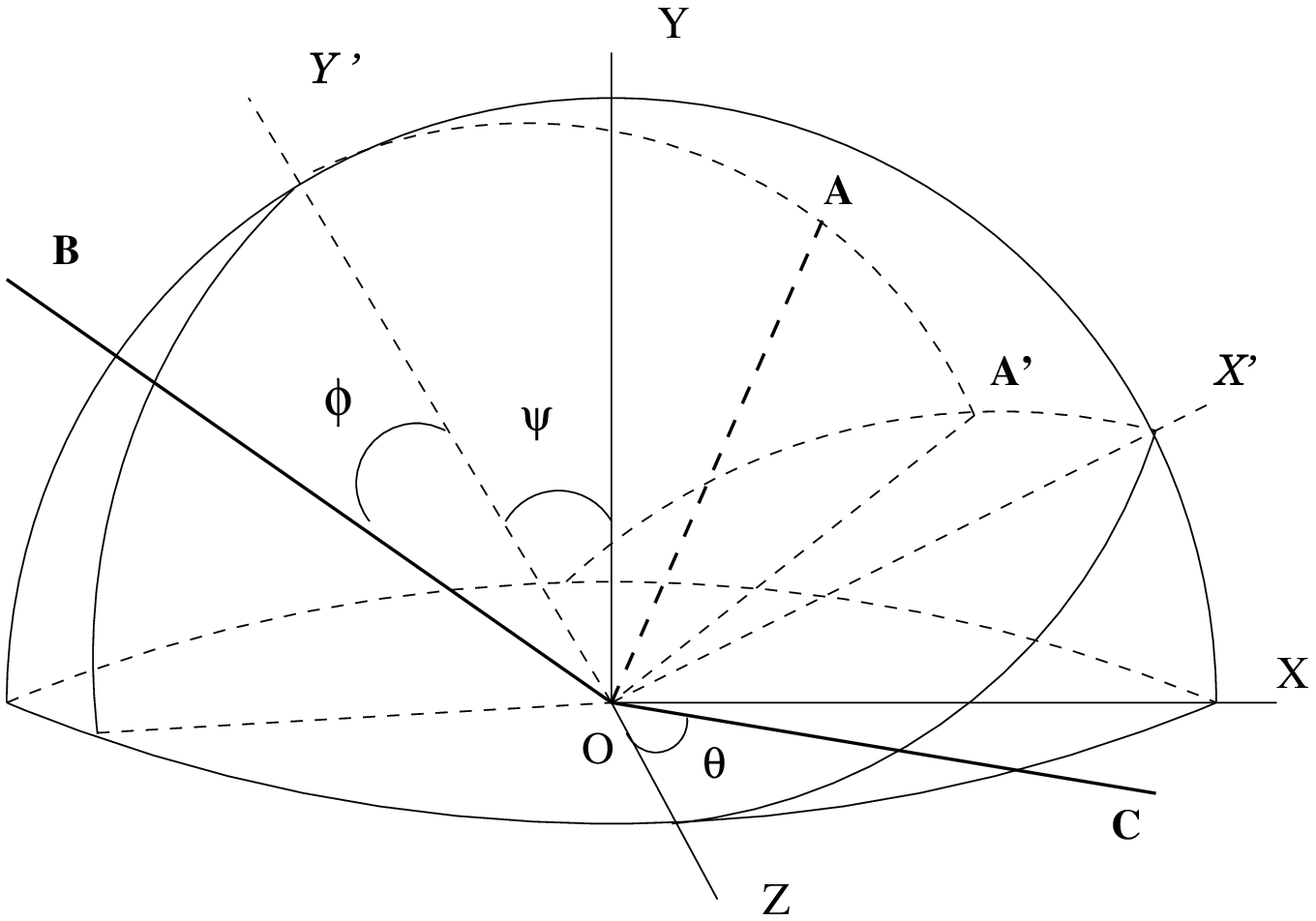}}
\caption{Rotary Motions of the Human Eye}
\end{figure}

\newpage
\begin{figure}
\centerline{
\epsfbox{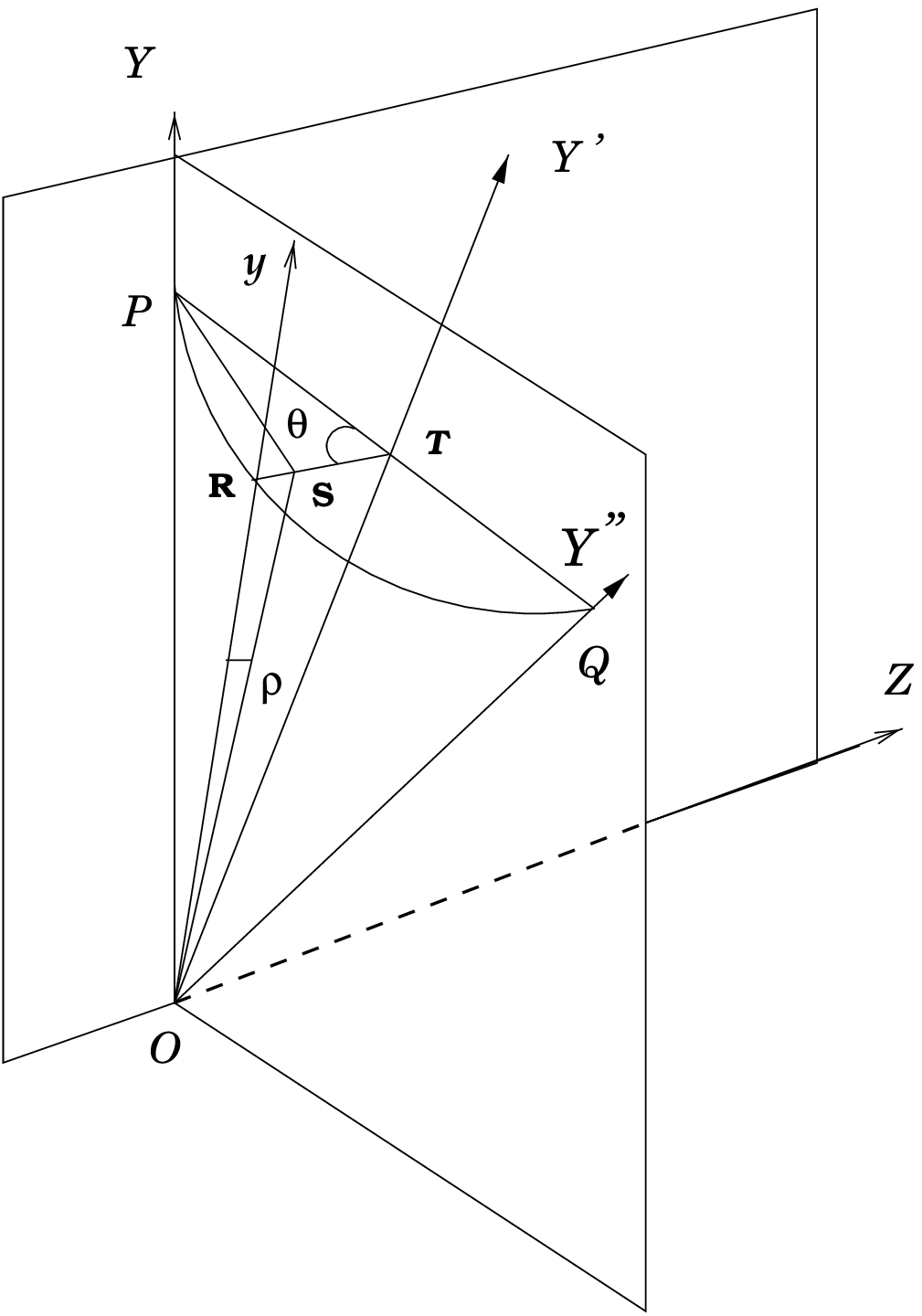}}
\caption{The Tilt of the Vertical}
\end{figure}


\begin{thebibliography}{9}

\bibitem{H} H. L. F. Helmholtz, {\it Treatise on Physiological Optics},
vol. 3, Translated by James P.C. Southall, Dover Publication, NY (1925) 
p.37-154.

\bibitem{S} J. P. C. Southall, {\it Introduction to Physiological
Optics}, Dover Publications, NY (1961) p.175.

\bibitem{He} D. Heller, {\it On the History of Eye Movement Recording}
in {\it Eye Movement Research, Physiological and Psychological Aspects},
G. Luer, U. Lass, J. Shallo-Hoffmann, eds., vol.2, C.J. Hogrefe, Toronto
(1988) p. 36.

\bibitem{L} H. Lamb, The Kinematics of the Eye, {\it Phil.Mag.} 
$6^{th}$ Series, {\bf 38} (1919) 685. 

\bibitem{K} K. Hepp, On Listing's Law, {\it Commun. Math. Phys.} 
{\bf 132} (1990) 285.

\bibitem{Ha} T. Haslwanter, Kinematics of Three-Dimensional Eye
Rotations, {\it Vision Research} {\bf 35} (1995) 1727.

\bibitem{DE} W. S. Duke-Elder, {\it Textbook of Ophthalmology}, 
vol. 1, The C.V. Mosby Co. St. Louis (1940) Chapter III.

\bibitem{O} J. van Opstal, Representation of Eye Position in Three
Dimensions. In Multisensory Control of Movement, A. Berthoz, Ed.,
Oxford, Oxford University Press (1993) p.27-41

\bibitem{Scar} J. B. Scarborough, {\it The Gyroscope, Theory and
Applications}, Interscience Publishers, London (1958) p.33.

\bibitem{C} R. H. S. Carpenter, {\it Movements of the Eye}, Pion, 
London (1977) p. 116-127.
 
\bibitem{D} H. Davson, {\it The Physiology of the Eye}, Pergamon Press, 
NY (1990) p. 648-656.

\bibitem{Ho} I. P. Howard, {\it Human Visual Orientation}, John Wiley,
NY (1982) p. 180-186.

\bibitem{A} M. Alpern, {\it The Kinematics of the Eye}, in {\it
The Eye} (2nd edition) vol. III, H. Davson, ed., Academic Press, 
NY (1969). 

\end{thebibliography}
\end{document}